\documentstyle[12pt]{article}
\begin{document}
\pagenumbering{arabic}
{\bf \quad} \\ [1cm]
\begin{flushright}
{\bf ISU-IAP.Th93-04,\ Irkutsk}
\end{flushright}
\ \\[1cm]

\begin{center}
{\Large \bf Is the scalar meson seen in CELLO data on
$\gamma\gamma\rightarrow\pi^+\pi^-$ ?  }\\[2cm]
{\Large A.E.Kaloshin and V.M.Persikov  }
\end{center}

\begin{center}
\sl Institute of Applied Physics, Irkutsk State University,\\
664003, Irkutsk, Russia.\ \ kaloshin@physdep.irkutsk.su\\[1.5cm]
\end{center}

\begin{center}
{\large \underline{ Abstract}.}
\end{center}

We analyze the CELLO angular distributions
$\gamma\gamma\rightarrow\pi^+\pi^-$ with the unitary model \cite{KS-86}
for helicity 2 amplitude. In contrast to previous analysis \cite{CELLO}
we do not see any QED damping. The obtained S--wave does not contradict
to low--energy theorem and demonstrates more clealy the resonance--like
behaviour near 1.3 Gev.
\newpage
\begin{center}
{\large 1. Introduction.}
\end{center}

In last two years first information has been
appeared on the angular distributions of
$\gamma\gamma\rightarrow\pi^+\pi^-$ from MARK-II \cite{M2} and
CELLO \cite{CELLO} detectors. It allows in principal to extract the
S--wave contribution on the background of dominant D--wave, where
exists well seen resonance $f_2 (1270)$. The reason of  special
interest to S--wave is related with the spectrum and
structure of mysterious scalar resonances I=0  (see, for instance,
review \cite{ADS}), which were not observed reliably in two--photon
processes up to now. The angular distribution analysis of CELLO data
( together with MARK-II data ) has been performed in \cite{CELLO,Harjes}.
A large S--wave cross--section has been found which was inerpreted
as  production
of a very broad resonance with mass $\sim 1100 $ Mev ( these data may be
seen at Fig.3 ). These results were discussed later from point of
view of resonance interpretation and scalar mesons spectrum. But close
inspection shows that the analysis \cite{CELLO,Harjes}
contains few very unexpected statements:

A) The data \cite{M2,CELLO} both on integral cross--section and
on angular distributions can not be described in  the simplest form
"Born term (QED) + Breit--Wigner $f_2 (1270)$".

B) CELLO and MARK-II data need the significant damping of QED
contribution. It was taken into account in analysis by means of factor
$1 /(1 + a \sqrt{s})^2$ and it was obtained $a \sim 2 \ Gev^{-1}$,
i.e. at \ $s = 1 \ Gev^2  $ \ the QED term is supressed in a few times.

C) The founded value of S--wave cross--section is very large: \\
$\sigma^S (\gamma \gamma \rightarrow \pi^+ \pi^- , \mid cos \Theta \mid
< 0.6 )\sim 60 - 80 $ \ nb  \ at \ $\sqrt{s} = 0.8 - 1.0 $ \ Gev.\\
\ \

Why these results are looking as unexpected ?
\begin{itemize}
\item The reasons of cross--section deviation from QED are the internal
structure of $\pi$--meson and final state interaction
and that's more these effects can not be separeted from each other.
In the vicinity of point $s=0, t=\mu^2$ it may be taken into account
in the low--energy theorem by introduction of structure constants --
pion polarizabilities. In the unitary models ( see,for instance,
\cite{KS-86} ) one never see so big deviation from QED as in \cite{CELLO}.
Note that in the final state interaction model the introducing of
some additional "QED damping factor " is rather inconsistent becouse
of double accounting of the same effects.
It may be shown also \cite{Terent'ev} in framework of
PCAC the absence of any formfactor in vertex
$\pi - \gamma -\pi$ if  $\gamma$ and one of $\pi$'s are on the mass shell.
As a whole we can say that so big deviation from QED is uncompatible
with low--energy theorem at the resonable values of
$(\alpha + \beta)^{\pi}$.
\footnote{\small In the helicity 2 amplitude the role of structure
constant plays the sum of polarizabilities $(\alpha + \beta)$
and in helicity 0 the difference $(\alpha - \beta)$.}
\item As for the value of S--wave cross--section
$\gamma\gamma\rightarrow\pi^+\pi^-$ CELLO, it is also uncompatible
with standard "final state interaction"  picture at reasonable values
of $(\alpha - \beta)^{\pi}$. To see it one can look at the predictions
\cite{KS-86}, where the typical value is
$\sigma^{++}(\gamma\gamma\rightarrow\pi^+\pi^-) \sim $\ 10 \ nb
near 1 Gev. Note that the recent analysis of experimental data
$\gamma\gamma\rightarrow\pi\pi$ \ \cite{KS-93} confirms it.
Extracting the values of pion polarizabilities from near--threshold data
and extrapolating cross--sections to the region $\sqrt{s} > 0.8 $\ Gev
one can obtain the cross--section value $\sim$\ 10 \ nb, but never
60--80 \ nb.
\item It is interesting that CELLO data on the integral cross--section
(i.e.before the separation on partial waves )
$\gamma\gamma\rightarrow\pi^+\pi^-$ in region  $\sqrt{s} > 0.8 $\ Gev
is compartible "by eye" with standard description \cite{KS-86},
though the amplitude \cite{KS-86} has another partial composition
as compared with \cite{CELLO}.\\[0.1cm]
\end{itemize}

These facts make to look more
carefully on the method of obtaining of these results.

In \cite{Harjes} an attempt was made to perform the model--independent
partial--wave analysis of CELLO data. But results are rather uncertain
becouse of uncomlete detector angle and presence of highest partial
waves. So the above--mentioned results  are obtained by another
and more model--dependent method. The idea is very simple and attractive:
let us fix the parametrization of $T_{+-}$ amplitude ( the main feature
here is the  well seen resonance $f_2(1270)$, interfereing
with smooth background ) and add an arbitrary S--wave contribution.
In such way the angular distributions at fixed s  allow to obtain
the S--wave cross--section , which we are interested, much more reliably .
In simbolic form:
\begin{equation}
T_{+-} = T_{+-}^{Born} + B.-W.\ f_2(1270)
\label{Symb}
\end{equation}

Our main statement is: all the unexpected results of CELLO  A) --- C)
are based on the specific suggestion on the form of Breit--Wigner
contribution in (\ref{Symb}). But in fact here does not exist big
freedom in description which is fixed by unitarity and analiticity
constrains ( plus low--energy theorem constrains ) \\ [0.1cm]

\begin{center}
{\large 2. On the form of helicity 2 amplitude.}
\end{center}

We want to compare the helicity 2 amplitude \cite{KS-86}
with one used in CELLO analysis  \cite{CELLO}, so it is necessary to
introduce some notations.
Let us introduce standard helicity amplitudes
$\gamma\gamma\rightarrow\pi\pi$ in CMS.
The cross--section $\gamma\gamma\rightarrow\pi^+\pi^-$ is defined as:
\begin{equation}
\frac{d \sigma}{d cos \Theta} = \frac{\rho(s)}{64 \pi s}  \
\{\ \mid T_{++}\mid ^2 + \mid T_{+-}\mid ^2  \} \\[0.5cm]
\end{equation}

Here $\rho(s) = (1-4\mu^2/s)^{1/2}$ ,\ $\mu= m_{\pi}$.
Let us pass over to reduced helicity amplitudes m which are free from
kinematical zeroes and singularities  \cite{AG}.
\begin{equation}
T_{++} = s M_{++}, \ \ \ \ \ T_{+-} = (tu-\mu^4) M_{+-}
\label{Reduc}
\end{equation}
The kinematical factors in (\ref{Reduc}) are originated from
angular momentum conservation and that is a very general constrain
on amplitudes.
We prefer to put on this constrain from the begining and to work
with the reduced helicity amplitudes m . By the way it allows to control
easy the low--energy theorem. If to separate Born term ( QED ) in
amplitudes, they have the following structure \cite{KS-86}:

\begin{equation}
M_{\Lambda}(s,t) = M_{\Lambda}^{Born}(s,t) + M_{\Lambda}^{C}(s,t)
\label{LE}
\end{equation}

$M_{\Lambda}^{C}(s,t)$ must be constant at the point $s=0, t=\mu^2$
and is related with pion polarizabilities.

Let us consider lowest partial wave with I=0 . The corresponding formula
from \cite{KS-86} is :
\footnote{To avoid misunderstandings: in \cite{CELLO,Harjes}
other amplitudes are used, though notations may coinside.
\cite{CELLO,Harjes} contains also mistake in isotopics ( sign )
which is unessential numerically. }
\begin{equation}
 m^{0,2} = [\ M_{+-}\ ]^{I=0,J=2} = V^0(s) + \frac{1}{D_f(s)}
[\ C^0+H^0_V(s)\ ]
\label{Our_amp} \\[0.5cm]
\end{equation}

Here $D^{-1}_f (s)$ is propagator of $f_2(1270)$ with the finite
width corrections . It is close numerically to simplest expression:
\begin{equation}
D_f(s) \approx m^2_f - s - i m_f \Gamma_f(s)
\end{equation}
$D_f(s)$ is apparently related with Omnes function:  \
$\Omega^0(s) = D_f(0)/ D_f(s)$.
$V^0$ in (\ref{Our_amp}) is defined mainly by QED term ,
resonance exchanges are relatively small.
\begin{equation}
V_0 \approx \frac{2}{\sqrt{3}} g^{\pi}(s) = \frac{2}{\sqrt{3}}
[\ M_{+-}^{Born}(s,t) \ ]^{J=2} \\[0.5cm]
\end{equation}

The $H_V^0(s)$  function is rescattering contribution
\begin{equation}
{\bf H^{0}_{V}(s)}=\frac{s}{\pi}\int \frac{ds'}{s'(s'-s)}
m_f \Gamma_f(s') V^{0}(s')
\label{Resc}
\end{equation}
One can rewrite (\ref{Our_amp}) in form close to used in \cite{CELLO}.

\begin{equation}
 m^{0,2} = \frac{R(s)}{D_f(s)} + V^0 \ \frac{Re D_f}{D_f} \approx \
\frac{R(s)}{m_f^2-s-im_f \Gamma_f(s)} +
V^0 \frac{m_f^2-s}{m_f^2-s-im_f \Gamma_f(s)} \\[0.5cm]
\label{Our_simp}
\end{equation}

where $R(s)=C^0+Re H_V^0(s)$. At resonance point R defines the decay
width $f_2 \rightarrow \gamma \gamma $ .
\begin{equation}
 R(m_f^2)=\frac{\sqrt{3} \cdot 2^5 \cdot 5\pi}{m_f^3}
\sqrt{\Gamma_f \cdot \Gamma(f \rightarrow \gamma \gamma) \cdot
BR(f \rightarrow \pi \pi)}
\end{equation}
The function R(s) in model \cite{KS-86} may be seen at Fig.2 .

If to compare (\ref{Our_simp}) with correspondent formula
from \cite{CELLO}, one can see that the only serious difference
consists in behaviour of  R(s). In the model \cite{KS-86} it is fixed
by D--wave phase shift and is defined by (\ref{Resc}).
In  \cite{CELLO} the following ansatz was used to fix the
R(s) form. The Breit--Wigner contribution to cross--section was chosen
as :

\begin{equation}
\frac{d \sigma^{BW}}{d \mid cos \Theta \mid} = 40 \pi \cdot
\frac{m_f^2}{s} \ \cdot
\frac{\Gamma_{\gamma \gamma} \ \Gamma_f(s) \ BR(f \rightarrow \pi^+ \pi^-)}
{\mid m_f^2 - s -i m_f \Gamma_f(s) \mid^2} \cdot
\mid Y_{22}(cos \Theta) \mid^2 \\[0.5cm]
\label{Ansatz}
\end{equation}

Here $\Gamma_f(s)$ is the same in nominator and denominator.
That is true for narrow resonance, when the s dependence is
unessential, but for rather broad $f_2$, interfering with background,
such choice of s--dependence in nominator looks as absolutely
arbitrary and unjustified.
\footnote{\small Similar ansatz evidently does not allow to describe
the $\rho$--meson in pion formfactor. }
$\Gamma_f(s)$ was taken as:
\begin{equation}
\Gamma_f(s)= \Gamma_f \cdot \left [ \ \frac{q(s)}{q(m_f^2)} \ \right ]^5
\cdot
\frac{m_f}{\sqrt{s}} \cdot f^2(s)
\end{equation}
Here f(s) is so called centrifugal barrier factor.
Returning from (\ref{Ansatz}) to lowest partial  wave
(\ref{Our_simp}), one can see that R(s) is of the form:
\footnote { \small In fact in analysis \cite{CELLO} was used \
f=1 , \ introducing of any decreasing with s factor  will make
the situation even worse, as noted in  \cite{CELLO}.}
\begin{equation}
R^H(s) = \frac{const}{s} \ f(s)
\label{R_H}
\end{equation}
First of all it is seen that such R(s) breaks the low--theorem
requirements (\ref{LE}) becouse of pole at  s=0 .
But for data analysis the more important is a quick increasing
of R(s)  below resonance. Becouse of interterence with background
it leads to large  D--wave cross--section exceeding the experimental one.
After it one needs to suppress the QED term by some manner and as result
it leads to above-- mentioned conclusions. \\[0.1cm]

\begin{center}
{\large 3. Analysis of CELLO data.}
\end{center}

Now let us try to analyze the CELLO data with using
of \cite{KS-86} amplitude. First of all consider the integral CELLO
cross--section (  $\mid cos \Theta \mid < 0.6$ )  and attempt to
describe it by this amplitude with one free parameter $C^0$.
Let us neglect for a moment by helicity 0 amplitude.
Fitting the region $\sqrt{s} < 1.5$ g\W , we shall obtain ( see Fig.1) :
\begin{equation}
C^0+Re H_V^0(m_f^2)= 0.314 \ \pm \ 0.003 \ Gev^{-2},
\Gamma( f_2 \rightarrow \gamma \gamma) = 3.46 \ \pm \ 0.05 \ Kev,
\end{equation}
\[
\chi^2 = 77.2 \ \   at  \  NDF=29
\]
The quality of description is not very good but we do not see some
conradictions. In any case there are no indications for large
S--wave croos--section .

Let us consider now the angular distributions of CELLO .
Note that function R(s) in  (\ref{Our_simp}) may be not fixed
beforehand but may be extracted from data together with scalar
contribution. Using the CELLO data ( 8 energy bins, 61 points )
we shall obtain the result depicted at Fig 2.
Summary $\chi^2$ is : $\chi^2 = 56.9 $  at  NDF=45 .
Such an anasysis shows that CELLO angular distributions prefer the more
smooth behaviour of R(s) as compared with (\ref{R_H}) and the smaller
S--wave cross--sections.
\footnote{ \small Much more evidently the smoothness of R(s)
is demonstrated by similar picture for MARK-II  data. About the  MARK-II
data see below.}
( The exception from common picture is only the first point with very
smooth angular distribution).

Let us repeat completely the partial--wave analysis
\cite{CELLO}
with using the model \cite{KS-86} instead of ansatz (\ref{R_H}).
In this way we extract $\Gamma(f_2 \rightarrow \gamma \gamma)$
and values of S--wave cross--section in 8 energy bins .
More exactly we proceed as follows: at first step we find $C^0$
as global parameter together with scalar contributions.
Then we  fix the best value of $C^0$ and independently find the S--wave
cross--sections in every energy bin.
\footnote{\small That corresponds obviously to procedure of
\cite{CELLO} which we are folowing here . }

The results of such an analysis are the following ( see  Fig 3 ).
\begin{equation}
\Gamma(f_2 \rightarrow \gamma \gamma) = 2.85 \ \pm 0.07 \ Kev,\ \
\chi^2 = 70.0 \ \ at \ \ NDF=53
\label{Our_val}
\end{equation}
Let us compare with similar result of  \cite{CELLO} ( which includes the
QED damping ) :
\begin{equation}
\Gamma(f_2 \rightarrow \gamma \gamma) = 2.58 \ \pm 0.13 \ Kev,\ \
\chi^2 = 81.4  \ \ at \ \ NDF=53
\end{equation}
Quality of description is approximately the same but but results are very
different as it is seen from Fig. 3.  \\[0.1cm]

\begin{center}
{\large 4. Conclusion.}
\end{center}

We showed that the source af all problems in partial--
wave analysis CELLO \cite{CELLO} is the rather arbitrary suggestion
(\ref{Ansatz}), (\ref{R_H}) on the energy dependence of "coupling
constant" R(s). The similar analysis of angular distribution with
using the model \cite{KS-86} leads to other conclusions:

a)  There is no necessity for QED damping to describe CELLO data both
on integral cross--section and angular distributions.
That is very naturally from our point of veiw.

b) The extracted  S--wave cross--section
$ \sigma^S ( \mid cos \Theta \mid < 0.6 ) $ is much less then obtained
in CELLO analysis and does not contradict to results of
threshold analysis  \cite{KS-93}.
We found that scalar conrtibution exists only near 1.2 -- 1.3 Gev,
out of this region the S--wave is compartible with zero. It gives
more serious indication for scalar meson production with mass about
1.3 Gev.

Few words about experiment of MARK-II
$\gamma \gamma \rightarrow \pi^+ \pi^-$  \cite{M2}.
In \cite{CELLO} in similar way were analyzed this data too, together
with CELLO or separetely. It was found practically complete coinsidence
of results for CELLO and MARK-II . We performed our analysis for
MARK-II data too. But results are differed:
formally MARK-II data need the negative S--wave cross--section in region of
$\sqrt{s} < 0.9$. The reason is that here the MARK-II data
are less then single QED contribution in contrast to CELLO.
Note that and previous experiments
$\gamma \gamma \rightarrow \pi^+ \pi^-$  do not agree with each other:
some of them are higher then QED, some are lower.
In the standard "final state interaction" models the helicity 2
cross--section
$\gamma \gamma \rightarrow \pi^+ \pi^-$ is higher than
electrodynamical one. In principal it is possibly to obtain
the cross--section lower then QED but only with rather high price.
It needs either the exotic behaviour of phase shift  \ $\delta^0_2 $\
between threshold and resonance or large high--energy contributions
( odderon in $\gamma \gamma \rightarrow \pi^0 \pi^0$ \cite{Zh} ?)
or marked dynamics in I=2  channel.

As for  MARK-II data one can say at the moment only that they
are not similar completely to
CELLO  ones at \  $\sqrt{s} < 0.9$ \ g\W. But the tendency is the same:
the abandon from ansatz (\ref{R_H})  leads to significant softening
of all contradictions.
We think here exist the principal physical question related with
hadron D--wave dynamics: is here something unusual or not.
It is possibly to try answer this question on the base of existing
experimental data.

We are grateful  to V.V.Serebryakov for discussions and participation
in initial stage of work. We thank J.Harjes for sending the experimental
information.

\begin{center} {\Large Figures captions.}
\end{center}
\begin{itemize}
\item {\bf Fig.1.} Results of CELLO cross--section fitting only by
$T_{+-}$ \cite{KS-86} in neglecting of $T_{++}$ amplitude.
The region $\sqrt{s} < 1.5$ g\W is fitting, as a result
$\Gamma( f_2 \rightarrow \gamma \gamma) = 3.46 \ \pm \ 0.05$ \ Kev,
$\chi^2 = 77.2$   at  NDF=29. For comparison the cross--section is shown
( H ) corresponding to (\ref{R_H}) with the same value $R(m_f^2)$.
\item {\bf Fig.2.} The obtained from CELLO angular
distributions the coupling "constant"  R \ (\ref{Our_simp}) \ and S--wave
contributions \ $\sigma^S$, $\mid cos \Theta\mid < 0.6 $ .
At Fig.2a first point is  $ R = - 2.0 \ \pm \ 1.2 \ Gev^{-2}$.
At Fig.2a the functions  R(s) are shown in the model \cite{KS-86}
and from \cite{CELLO}.
\item {\bf Fig.3.} The repetition of analysis \cite{CELLO} with using
of model \cite{KS-86} instead of ansatz (\ref{R_H}).
For comparison results of \cite{CELLO} are shown ( crosses).
The curve ( same as at \ Fig.1 \ ) \ demonstrates the scale of
integral CELLO cross--section.
\end{itemize}
\end{document}